\documentclass[floats,twocolumn,prl,aps]{revtex4}
\usepackage{latexsym}
\usepackage{graphicx}

\newcommand\beq{\begin{equation}}
\newcommand\eeq{\end{equation}}
\newcommand\bea{\begin{eqnarray}}
\newcommand\eea{\end{eqnarray}}
\newcommand{\nonum}{\nonumber}

\begin{document}
\title{\bf Quantum Phase Analysis of Extended Bose-Hubbard Model  
}

\author{\bf Manoranjan Kumar$^1$, Sujit Sarkar$^2$, S. Ramasesha$^1$}

\address{\it 
$^1$Solid State and Structural Chemistry Unit, Indian Institute of Science, Bangalore 5600 12, India,
$^2$Poornaprajna Institute of Scientific Research, 4 Sadashiva Nagar, Bangalore 5600 80, India.}    
\date{\today}

\begin{abstract}
We have obtained the quantum phase diagram of one dimensional extended Bose-Hubbard model 
using the density-matrix renormalization group and Abelian bosonization methods for different 
commensurabilities. We describe the nature of different quantum phases at the charge degeneracy point. 
We find a direct phase transition from Mott insulating phase to superconducting 
phase for integer band fillings of bosons. We predict explicitly the presence of two kinds of  
repulsive Luttinger liquid phases, apart from the charge density wave and 
superconducting phases for half-integer band fillings. Our study reveals that extended range interactions 
are necessary to get the correct phase boundary of an one-dimensional interacting bosons system.

\vskip .4 true cm
\noindent PACS numbers: ~74.78.Na 
, 74.81.Fa,
74.20.Mn

\end{abstract}
\maketitle
%\begin{multicols}{2}
\section{ 1. Introduction}
We present the quantum phase diagram (QPD) of extended Bose-Hubbard model (EBH) \cite{subir,MPA,white},
 for spinless bosons on a one-dimensional lattice. 
These bosons could represent Cooper pairs undergoing Josephson tunneling
between superconducting islands or Helium atoms moving on a substrate.
Here we are interested in the quantum phase transitions of the ground state.
 This transition is driven by quantum
fluctuations of the system which are controlled by the system parameters \cite{subir,sondhi}. 
There are a few experiments carried out on one-dimensional Josephson 
junction systems of different lengths under a magnetic field \cite{ havi1,zant,havi2,havi3}; a 
superconductor-Motts insulator transition is observed. In these systems, 
relevant particles are the Cooper
pairs which can be treated as  spinless charge bosons. Hence the Bose-Hubbard 
model and its variants are prominent candidates to study such systems.
 It is also revealed from experiments that the inter-particle interactions are 
of finite range \cite{white,havi1,zant,havi2,havi3}. We study an EBH model
that incorporates the nearest-neighbor (NN) and next-nearest-neighbor (NNN) 
repulsive interactions in addition to the on-site Hubbard repulsion.
 We also consider the NN and NNN hopping of bosons on the lattice. 
In what follows, we derive an analytical expression for Mott insulating
 gap at the mean-field limit but this alone is not sufficient to capture 
the effects of all interactions especially at the charge degeneracy point \cite{sar1}. 
We map our EBH model to a spin chain model, for strong on-site 
Coulomb repulsions. In this letter, we first present
the results of Abelian bosonization study of spin chain model,  
we then discuss the results of our Density Matrix 
Renormalization Group method (DMRG) \cite{white1} study of the EBH model.  

The model Hamiltonian representing, arrays of superconducting (SC) 
nano-island, here after refered to as superconducting quantum dots (SQD) is 
given by,

\bea
H & =&   -t_1 \sum_{i} (\hat b^{\dagger}_{i} \hat b_{i+1}  + h.c ) 
- t_2 \sum_{i} ( \hat b^{\dagger}_{i} \hat b_{i+2}  + h.c )  \nonum\\ 
& &  + \frac{U}{2} \sum_{i} \hat n_i( \hat n_i ~-~ 1) 
+ {V_1} \sum_{i} \hat n_{i} \hat n_{i+1} \nonum\\ 
& &+ V_2 \sum_{i} \hat n_i \hat n_{i+2}-\mu \sum_{i} \hat n_{i}
\eea

in the usual notation.

\section{ 2. Model Hamiltonians and Continuum Field Theoretical Study:}
 We recast our basic Hamiltonians in the spin 
language \cite{lar} to obtain;
%\begin{center}
$H_{J1}= -2~E_{J1} \sum_{i} ( S^+_{i} S^{-}_{i+1} + h.c)$,
$H_{J2}=-2~E_{J2} \sum_{i}(S^+_{i}S^{-}_{i+2} + h.c)$, 
%\end{center}.
%\begin{center}
$H_{E_{C0}}= \frac{E_{C0}}{2} \sum_{i} 
{(2 {S^Z}_{i} - h )^{2} }$,
$H_{E_{C1}}=4E_{Z1} \sum_{i} S^Z_{i}S^{Z}_{i+1}$,
$H_{E_{C2}}=4E_{Z2} \sum_{i} S^Z_{i}S^Z_{i+2}$.
%\end{center}
% see the following care fully 
Our total Hamiltonian is, 
\begin{equation}
H= H_{J1}+H_{J2}+H_{Ec0}+H_{Ec1}+H_{Ec2}
\end{equation}

Here $h=(N - 2n -1)/2$ and N tunes  
the system to a degeneracy point by means a gate voltage($eN \sim V_g$) where $eN$ is the 
average dot charge induced by the gate voltage.
The phase diagram is periodic in $N$ with period 2. 
Here we only consider a single slab of the phase diagram,
$0 \leq N \leq2$. Correspondence between the parameters
of the EBH model and the spin chain model is;  
 $\langle n \rangle$$t_1$ $\sim E_{J1}$,
 $\langle n \rangle$${t_2}$ $\sim E_{J_2}$, 
U $\sim E_{co}$, $V_1$ $\sim E_{z1}$, $V_2$ $\sim E_{z2}$ \cite{fazi}. 
When the ratio $\frac{E_{J1}}{E_{C0}} \rightarrow 0$, the SQD 
array is in the insulating state with gap 
$\sim {E_{C0}}$, it costs an energy $\sim E_{C0}$
to increase by one, the number of pairs on a given dot. 
Exceptions are the discrete points at $N=2n+1$, where a dot
 with charge $2ne$ or $2(n+1)e$ has the same energy because 
gate charge compensates the charges of extra Cooper pair in the dot.
At this degeneracy point, a small amount of Josephson coupling 
leads the system to the SC state. To analyze the 
phases explicitly near this degeneracy point, we recast 
the Hamiltonian of the spin chain model to a spinless fermion model
through Jordan-Wigner transformations and carry out a continuum 
field theoretical study of spinless fermion model \cite{gia1}.   
 Our continuum models, in terms of bosonic fields \cite{sar2}, are
\bea
H_1 & = & H_0 ~-~\frac{4 E_{Z11}}{{(2 \pi \alpha)}^2}
\int \cos( 4 \sqrt{K} {\phi } (x)~)~ dx~\nonum\\ 
~& & +\frac{2 h E_{C0}}{\pi} \int ~{{\partial}_x}{\phi}~dx.
\eea
\bea
H_2 & = & H_0 ~-~\frac{4E_{Z12}}{{(2 \pi {\alpha})}^2} 
\int \cos( 4 \sqrt{K} {\phi } (x)~)~ dx ~\nonum\\
~& & +\frac{2 h E_{C0}}{\pi} \int ~{{\partial}_x}{\phi}~dx.
\eea

$
H_0 ~=~ \frac{v}{2 \pi} ~\int dx ~[ 
{({{\partial}_x} \theta)}^2
+ {({{\partial}_x} \phi)}^2~],
$
 is the non-interacting part of the Hamiltonian (Eq.2).
The velocity, $v$, of low energy excitations is one of 
the Luttinger liquid (LL) parameters while $K$ (Eqs. 3 and 4) is the other.

$\phi$ field 
corresponds to the quantum fluctuations (bosonic) of spin and $\theta$ is 
the dual of $\phi$. They are related by ${\phi}_{R}=\theta-\phi$ and 
 ${\phi}_{L}=\theta+\phi$.
The effect of applied gate voltage on
the SQD appears as an effective magnetic field in the
spin representation of these Hamiltonians. 
In $H_1$, we consider the NN Josephson coupling, on-site and 
NN charging energies of SQD and also the NN charging
energy correction for the co-tunneling process. In $H_2$,
we consider all the interactions present in our Hamiltonian
 besides all terms of co-tunneling process \cite{lar,sar2}. Analytic
 expressions for $E_{Z11}$ and $E_{Z12}$ are 
($E_{Z1}-\frac{3 {E_{J1}}^2}{16 E_{C0}}$) and 
($E_{Z1}-\frac{3 {E_{J1}}^2}{16 E_{C0}}-\frac{3}{2} E_{Z2}$)
respectively.
\section{ 3. Numerical studies}
We use the DMRG method to numerically study the QPD of the Hamiltonian 
in Eq. 1. We employ the infinite DMRG algorithm keeping 128 
density matrix eigenvectors. The site boson Fock space is 
truncated to three and the number of sites in the chain is restricted 
to 128. Per site bosonic filling fractions 0.5 and 1.0 are studied. 
Accuracy of the method is checked by comparing the ground state energies, 
various correlation function and charge gap from DMRG studies with exact 
diagonalization studies of small systems. We have also reproduced the 
result of earlier DMRG calculations satisfactorily \cite{white}. The discarded density 
in the DMRG calculations is less than $10^{-14}$ in the charge density wave (CDW) phase
 at $\rho=0.5$ as well as in the Mott-insulating phase at $\rho=1.0$. 
However, the discarded density is slightly less than $10^{-10}$ in the SC phase.
For calculating the correlation function, we target only the 
ground state. To obtain the Berezinskii-Kosterlitz-Thouless (BKT) transition 
point, we plot the correlation function $<b_r^+b_0>$ vs $r$ on a log-log plot, 
here $r=0$ is the middle site in the chain. The slope of this plot 
in the region between $r=16$ and $r=36$ gives $-K/2$, where $K$ is the
Luttinger liquid (LL) parameter (Eqs. 3 and 4). The BKT-point corresponds to slope of -1 for 
$\rho=0.5$ and -0.25 for $\rho=1.0$.
%***************************************************************
\section{ 4. Physical Analysis of QPDs.}
Here we analyze the QPD of Hamiltonian $H_1$ (Eq. 3). 
The analytical structure of this model is the
same as that of the XXZ Heisenberg chain in a magnetic field. Hence, 
LL parameter of $H_1$ can be calculated exactly by using the Bethe-ansatz
 techniques. Analytical expression for $K$ \cite{gia1} is of the form
\beq
K~=~ \frac{\pi}{\pi + 2 \sin^{-1} \Delta},
\eeq 
where $\Delta$ is the XXZ anisotropy parameter of the Heisenberg chain.
%During our analysis, we will use two different $\Delta$,
%${\Delta}_{1}$ (where $E_{Z11}~=~E_{Z1}$). 
%and ${\Delta}_{2}$.
Our analysis will use the two limits of the parameters $\Delta$, namely 
$\Delta_1$ and  $\Delta_2$; the analytical expressions 
for these are 
$\frac{2 E_{Z1}}{E_{J1}}$
and  
($\frac{2 E_{Z1}}{E_{J1}} - \frac{3 E_{J1}}{8 E_{C0}}$)  
respectively.

We calculate the analytical expression for the 
LL parameter for Hamiltonian $H_2$ (Eq. 4) as
\beq
K~=~\sqrt{\bigg[\frac{ E_{J1}~-~ \frac{32}{\pi} E_{J1} E_{Z2} }
{E_{J1}~+~\frac{2}{\pi} (4 E_{Z1}~ -~\frac{3 {E_{J1}}^2 }{4 E_{C0}})}\bigg]}
\eeq  

\begin{figure} 
\includegraphics[scale=0.25,height=5.2cm,width=9.0cm,angle=0]{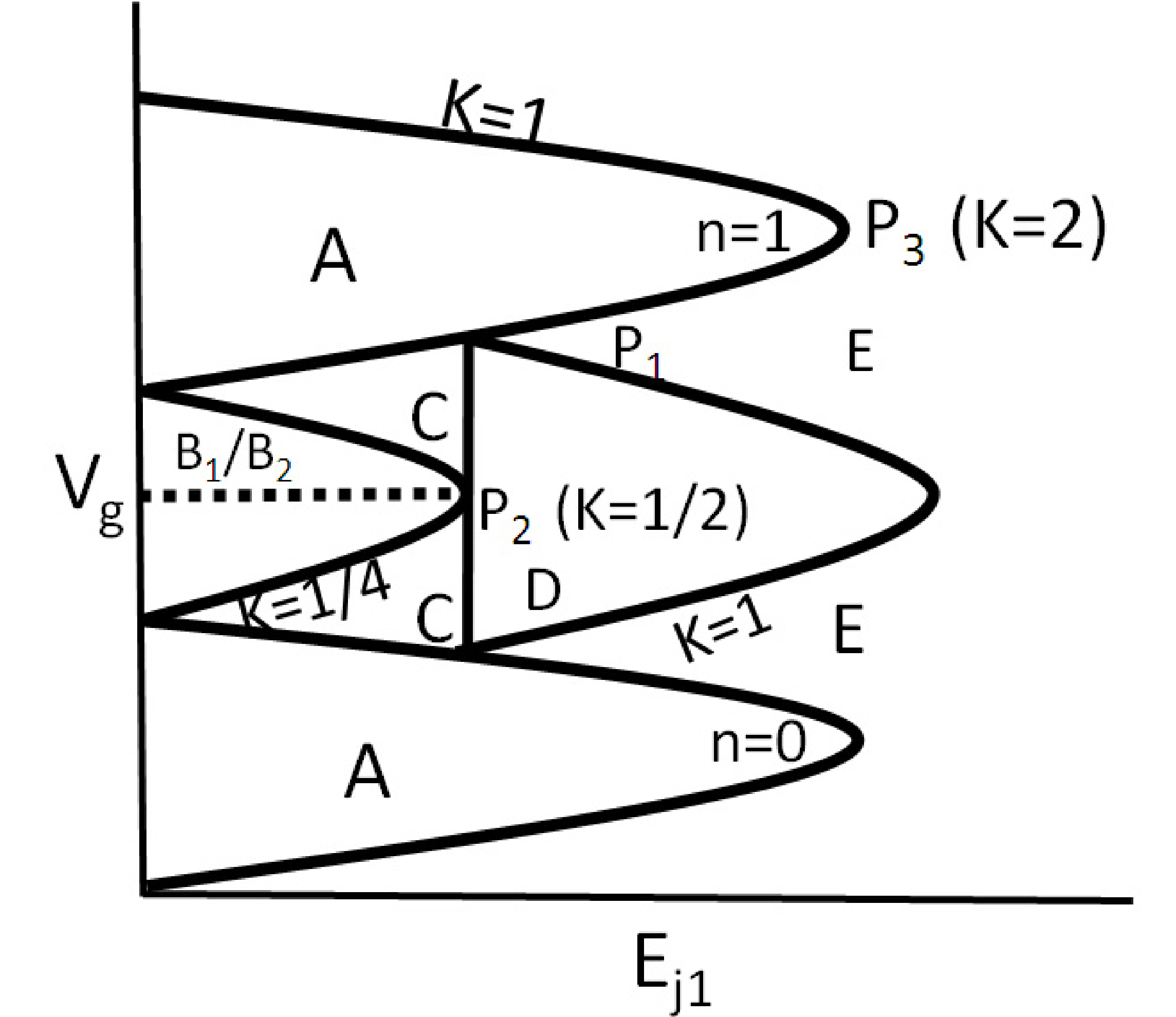} 

\caption{Quantum phase diagram ($E_{J1}$ vs. $V_g$) of
the SQD array. 
A is the Mott insulating phase, $B_1$ is the CDW phase, 
$B_2$ is the Dimer-order density wave, C is first kind of Repulsive Luttinger
 liquid (RL1), D is second kind of Repulsive Luttinger liquid (RL2) and  E is 
the SC phase. $P_1$ and $P_2$ are the multi-critical points (please see text).
 $K$ is one at the Luttinger liquid and SC phase boundary and
the Mott phase and SC phase boundary. $K =1/4$ at the phase boundary between 
RL1 and CDW state. $K=1/2$ and 2 at $P_2$ (charge degeneracy) and $P_3$ (particle-hole symmetry) points respectively.}
\label{fig1}
\end{figure}
\begin{figure} 
\includegraphics[height=6.2cm,width=9.0cm,angle=0]{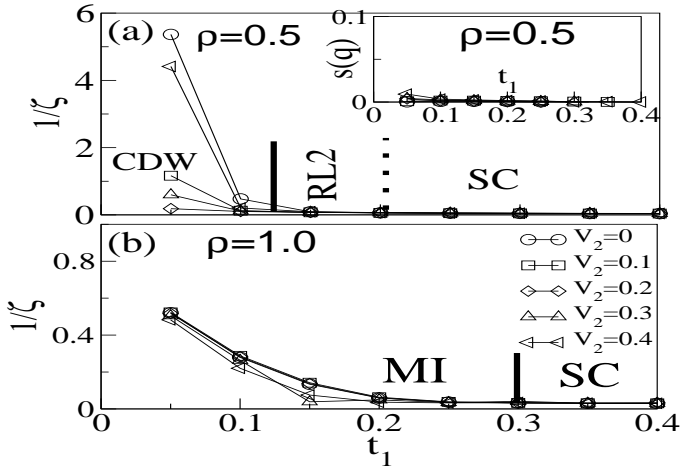} 
\caption{Phase diagram of $1/{\xi}$ vs. $t_1$ for (a) $ \rho =0.5 $ and (b) $ \rho =1.0 $  boson density 
for different values of $V_2$ $(\xi^2=\sum_{r}{r^2<b^\dagger_r b_{0}>}/ \sum_{r}{<b^\dagger_r b_{0}>})$.
We predict three quantum phases: CDW, RL2 and SC phase for $ \rho =0.5 $ but there is no evidence of RL2 
phase for $ \rho =1.0 $. Inset in (a) show structure factors for 
$ \rho =0.5 $ .}

\label{fig2}
\end{figure}

\begin{figure}
\includegraphics[height=4.0cm,width=6.5cm,angle=0]{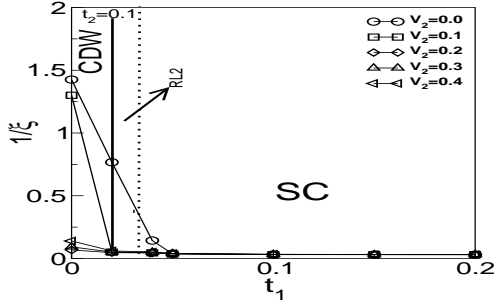}
\caption{Phase diagram in ($t_1$, $1/{\xi}$) plane for boson density of 0.5.
 We predict effect of finite value of $t_2$ on two phase boundaries (CDW-RL2, 
RL2-SC).}

\label{fig3}
\end{figure}

\noindent In the limit $\Delta={\Delta}_1$,
for $E_{J1} < 2 E_{Z1}$ and
relatively small field, the anti-ferromagnetic  Ising interaction
dominates the physics of anisotropic Heisenberg chain. When the 
field is large, i.e., for large applied gate voltage, 
the chain state is in a ferromagnetic state.
In the language of interacting bosons, the N\'{e}el phase is the 
commensurate CDW phase with period 2.
In Fig. \ref{fig1}, this phase region is described by region $B_1$. 
The ferromagnetic state is the Mott insulating state, this is the
phase A of our quantum phase diagram (Fig. \ref{fig1}).
The emergence of two LL phases can be attributed to 
following reason: RL1 ($K<1/2$) occurs due to
commensurate-incommensurate transition while  RL2 ($K>1/2$) 
occurs due to the critically of the XY model.  The physical 
significance of RL1 (region C in Fig. \ref{fig1}) phase is that the coupling term is relevant and 
 the applied magnetic field (the applied gate voltage), breaks 
the gapped phase. In the RL2 phase (region D in Fig. \ref{fig1}) 
none of the coupling terms is 
relevant due to large values of $K$ ($>1/2$). 
These phase regions, are similar for the two different limits, 
$\Delta = {\Delta}_1$ and $\Delta = {\Delta}_2$.
The analysis of the physics of two RL phases is entirely new. \\

In Fig. \ref{fig2},  we present the DMRG study of EBH model
(Eq. 1) for boson densities, $\rho$, of $0.5$ (Fig. \ref{fig2}a) and $1.0$ (Fig. \ref{fig2}b).
We predict three quantum phases, CDW, RL2 and SC for
 $\rho=0.5$. There is no evidences of LL phase for
integer boson density. So our field theoretical analysis is consistent
with DMRG studies. The structure factor in inset of Fig. \ref{fig2}a for $\rho=0.5$ precludes 
CDW phase for this parameter range. There is also no evidence for RL1 
phase; one obtains the RL1 phase 
for higher values of $\mu$ (gate voltage) in the CDW regime. We observe that
the boundary between CDW and RL2 phases as well as RL2-SC phase boundary 
shift slightly for higher values of ${V_2}$. Fig. \ref{fig3} shows the effect of NNN hopping on the phase 
diagram of Fig.\ref{fig2}. We note that the RL2 phase and CDW region are squeezed on 
introducing the NNN hopping while expanding the region of SC phase. 
In Figs. \ref{fig2} and \ref{fig3}, we have  presented the
explicit parameter dependence of QPD which is not possible to obtain from the
Abelian bosonization study.\\

If we consider only the interaction terms of the Hamiltonian
$H_1$, for both ${\Delta}_1$ and ${\Delta}_2$ limits we get the 
condition $E_{Z1} \leq
-\frac{E_{J1}}{2}$ for the boundary between RL1 and CDW.
This condition is unphysical; we know experimentally that 
 $E_{J1}$, $E_{J2}$, $E_{Z1}$ and $E_{Z2}$ are all
positive \cite{havi3,fazi}. So the interaction part of Hamiltonian 
$H_1$ alone is not sufficient to produce the whole phase
diagram of Fig. \ref{fig1}. It indicates that an extended interaction 
term is required to get the correct phase diagram.
%We want to discuss the phase boundary between RL2 and 
We focus phase boundary between RL2 and SC phase of $H_1$ for 
$\Delta={\Delta}_1$ and ${\Delta}_2$, similar analysis of LL parameter 
gives the condition $E_{Z1}<0$ which is again unphysical. This phase boundary also 
requires NNN hopping and NNN coulomb interactions.   
\begin{figure}
\includegraphics[height=5.2cm,width=7.0cm,angle=0]{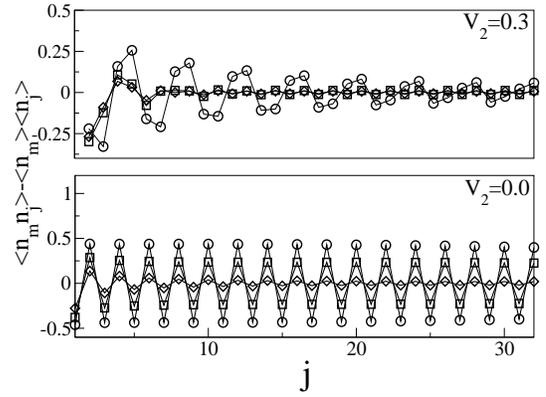}
\caption{Density-density correlation function for two values of
$V_2$ at $V_1=0.4$. $\circ-\circ ~t_1=0.05$, $\Box - \Box ~t_1=0.1$ 
and $\Diamond-\Diamond ~t_1=0.15$.}
\label{fig4}
\end{figure}

\begin{figure}
\includegraphics[height=6.2cm,width=8.0cm,angle=0]{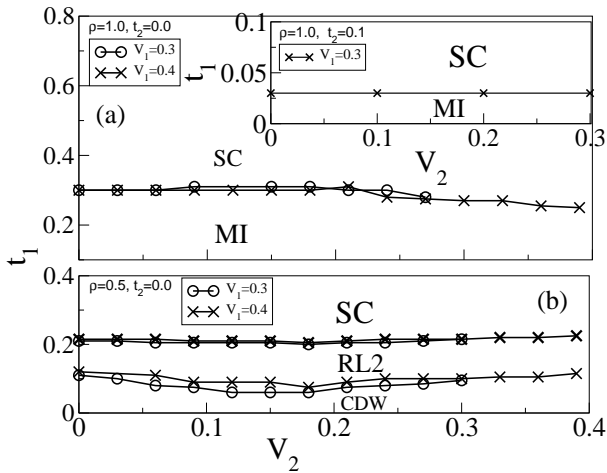}
\caption{$t_1$ vs. $V_2$ phase diagram for (a) $\rho=1.0$ and (b) $\rho=0.5$  
(BKT) transition for two values of $V_1=0.4$ and $0.3$. Different points mark the different BKT points.
Inset in (a) shows effect of finite values of $t_2$.}
\label{fig5}
\end{figure}
Here we do the analysis of the Hamiltonian $H_2$, which extends the range of hopping 
as well interactions of $H_1$, to NNN. Analysis for  RL1 and CDW boundary, 
yields a quadratic equation in $E_{J1}$. The real roots of the quadratic 
equation do not impose negative sign for $E_{J1}, E_{J2}$, $E_{Z1}$ and $E_{Z2}$ unlike in the case with $H_1$.
 Therefore, we can find physical values of these parameters for which RL1 and CDW phase boundary can 
exist. Similar analysis for the phase boundary between RL2 and SC phase indicates that $H_2$ can 
also support such a phase boundary. $B_2$ phase of Fig. \ref{fig1} can be obtained only from the 
analysis of Hamiltonian $H_2$ when $E_{z2}$ exceeds some critical value like in the Majumdar-Ghosh model \cite{mazum}.  

%So the presence of NNN
%Josephson coupling gives us a physical footing for the existence
%of this boundary. 

The above analysis is new and a consequence of extending the range of hopping and interactions. 
In Fig. \ref{fig1}, we find two multi-critical point, $P_1$, and $P_2$. Mott, RL1, RL2, and superconducting phases coincide at the 
point $P_1$; RL1, RL2, and $B_1$ or $B_2$ coincide at the $P_2$ point. In our QPD,
we obtained multi-critical points not only for the presence
of different kinds of interactions in the SQD but also on application of external gate 
voltage on the SQD. In our QPD, the class of transition is changing as one moves away 
from the tip of the lobe in the Fig. \ref{fig1}. Our analysis  allows to extend the 
classification of two types of Mott transition in the context of SQD. 
At the tip of each lobe, the system at fixed density is driven from the SC phase to 
insulating phase by changing the interaction parameter and this is  Mott-U 
(Berezinskii-Kosterlitz-Thouless) transition. Away from the tip of the lobe, a change in 
the gate voltage drives the system from a SC phase to an insulating phase and this is
 the Mott-$\delta$ transition.\\ 

Fig. \ref{fig4} shows that density-density correlation function $F(j)=(<n_mn_j>-<n_j><n_m>)$ 
 from the middle site of the chain, $m$. We study  $F(j)$ for  different values of $V_2$ and 
$t_2$. We observe that as $V_2$ increases the system crosses over from the $2K_F$ CDW 
state to $4K_F$ CDW state. Increasing $t_2$ leads to decrease in CDW amplitude. The prediction of $4 K_F$ CDW state 
is entirely new for this Hamiltonian.

In Fig. \ref{fig5}, we present the DMRG analysis of BKT transition for $\rho=0.5$ and $\rho=1.0$ for different values of $V_1$. 
We observe that  for $\rho=0.5$, SC phase occurs for small values of $t_1$ compared to $\rho=1.0$.
 For $\rho=0.5$, CDW phase and RL2 
boundary appears at $t_1=0.11 \pm 0.03$ for $V_1=0.3$ and 0.4 for different values of $V_2$. 
We also note that $t_2=0.1$ shifts the MI-SC boundary to a much smaller value of $t_1$($<0.03$). 

In summery, we have presented the phase diagram of EBH model. Description of repulsive luttinger liquid 
phase regions and the $4 K_F$ CDW states for higher values of $V_2$ is entirely new.    

Acknowledgement: MK thanks UGC, India for felloship, SS thanks dept. 
of Physics, IISc for facilities extended. This work was supported in part by 
a grant from DST (No.SR/S2/CMP-24/2003), India.

\end{document}